\documentclass[showpacs,pra]{revtex4}
\usepackage{graphicx}
\begin{document}
\title{Reflection and Splitting of Channel-Guided Solitons in  Two-dimensional Nonlinear Schr\"odinger Equation}
\author{Hidetsugu Sakaguchi and Yusuke Kageyama}
\affiliation{Department of Applied Science for Electronics and Materials,
Interdisciplinary Graduate School of Engineering Sciences, Kyushu
University, Kasuga, Fukuoka 816-8580, Japan
}
\begin{abstract}
Solitons confined in a channel are studied in the two-dimensional nonlinear Schr\"odinger equation. When a channel branches into two channels, a soliton is split into two solitons, if the initial kinetic energy exceeds a critical value.  The branching point works as a pulse splitter.  If it is below the critical value, the soliton is reflected. The critical kinetic energy for splitting is evaluated by a variational method.  The variational method can be applied in the design of channel systems with reduced reflection. 
\end{abstract}
\maketitle

The one-dimensional nonlinear Schr\"odinger equation is a typical soliton equation. Optical solitons in optical fibers are described by the nonlinear Schr\"odinger equation.\cite{rf:1} The Bose-Einstein condensates (BECs) can be described by the Gross-Pitaevskii equation, which is equivalent to the nonlinear Schr\"odinger equation with a potential term.\cite{rf:2} Bright and dark solitons have been observed in several experiments of the Bose-Einstein condensates.\cite{rf:3,rf:4} 
The one-dimensional nonlinear Schr\"odinger equation has been sufficiently studied, but solitons in two- or three-dimensional nonlinear Schr\"odinger equations are not so intensively studied. We studied solitons in guided channels in the two-dimensional nonlinear Schr\"odinger equation.\cite{rf:5} In optical systems, refraction-index-guiding channels are used to confine optical pulses.  Matter-wave solitons in BECs are created in cigar-shaped traps, which work as guiding channels.  Solitons can propagate along the guiding channel with an arbitrary velocity if the guiding channel is uniform and the norm of the soliton is below a critical value for the collapse. 

In this paper, we study the motion of solitons in inhomogeneous guiding channels. In particular, we study the motion of a soliton when a channel branches into two channels. Such a branching point is important as a pulse splitter in optical systems and a beam splitter for atomic waves.\cite{rf:6} 
There has been no direct numerical simulation of the splitting of two-dimensional solitons in the branching waveguide up to now. 
We will perform a variational approximation using the Lagrangian to understand the numerical results.  The approximation is a generalization applied to the one-dimensional nonlinear Schr\"odinger equation under an external potential.\cite{rf:7}

The model equation is written as 
\begin{equation}
i\frac{\partial \phi}{\partial t}=-\frac{1}{2}\nabla^2\phi-|\phi|^2\phi+U(x,y)\phi,
\end{equation}
where $U(x,y)$ denotes the potential, which represents a channel. For a straight channel, $U(x,y)=-U_0$ for $-x_0\le x  \le x_0$ and $U(x,y)=0$ for other 
regions. The width of the channel is denoted as $2x_0$ and the depth of the potential is denoted as $U_0$. 
A stationary solution to eq.~(1) can be numerically obtained from the time evolution of the Ginzburg-Landau-type equation with a feedback term:
\begin{eqnarray}
\frac{\partial \phi}{\partial t}&=&\frac{1}{2}\nabla^2\phi+|\phi|^2\phi-U(x,y)\phi+\mu\phi,\nonumber\\
\frac{d\mu}{dt}&=&\alpha(N_0-N),
\end{eqnarray}
where $\mu$ is an additional variable corresponding to the chemical potential, $N=\int\int |\phi|^2dxdy$ is the total norm, and $N_0$ is the target value of the total norm. By a long-time evolution of eq.~(2), a two-dimensional soliton with norm $N_0$ is obtained as an attractor of the dynamical system. 
Figure 1(a) displays a 3D plot of $\phi$ for $N_0=5, U_0=5$, and $x_0=3$. 
Figure 1(b) shows the profile $|\phi(x_p,y)|$ in the section of $x=x_p=0$, and Fig.~1(c) shows $|\phi(x,y_p)|$ in the section of $y=y_p$, where $(x_p,y_p)$ is the peak position of the soliton solution. 
\begin{figure}[tbp]
\begin{center}
\includegraphics[height=3.5cm]{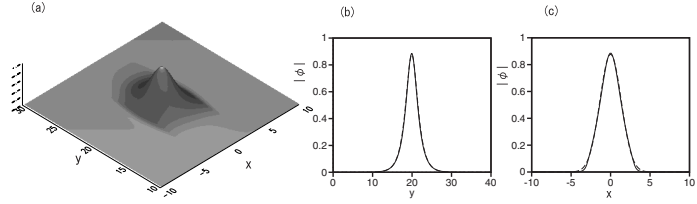}
\end{center}
\caption{ (a) 3D plot of $|\phi|$ at $x_0=3,U_0=5$, and $N_0=5$. (b) Profile of $|\phi|$ in the section of $x=0$. The dashed curve is an approximate solution obtained by the variational method,  which well overlaps the numerical result. (c) Profile of $|\phi|$ in the section of $y=y_p$, where $(0,y_p)$ is the peak position of $|\phi(x,y)|$ shown in (a). The dashed curve is the approximate solution obtained by the variational method.}
\label{f1}
\end{figure}

The Lagrangian corresponding to the nonlinear Schr\"odinger equation (1) is expressed as 
\begin{equation}
L=\int\int [(1/2)\{i(\partial \phi/\partial t)\phi^*-i(\partial \phi^*/\partial t)\phi\}-(1/2)|\nabla\phi|^2+1/2|\phi|^4-U(x,y)|\phi|^2]dxdy.
\end{equation}
If $\phi$ is approximated as $\phi=A{\rm sech}(y/a)\exp\{-x^2/(2b^2)\}\exp(-i\mu t)$, the Lagrangian is calculated as
\begin{equation}
L_{e}=N\{-\frac{1}{4b^2}-\frac{1}{6a^2}+\frac{N}{6\sqrt{2\pi}ab}+U_0{\rm erf}(x_0/b)\},
\end{equation}
where $N=\int\int|\phi|^2dxdy=2A^2\sqrt{\pi}ab$ is the total norm and ${\rm erf}(x)=2/\sqrt{\pi}\int_0^{x}\exp(-z^2)dz$ is the error function.
The variational principle $\partial L_{e}/\partial a=\partial L_{e}/\partial b=0$  yields coupled equations for $a$ and $b$ as 
\begin{equation}
a=2\sqrt{2\pi}b/N,\;\;1/2-N^2/(24\pi)+2U_0/\sqrt{\pi}\exp(-x_0^2/b^2)(-x_0b)=0.
\end{equation}

\begin{figure}[bp]
\begin{center}
\includegraphics[height=3.5cm]{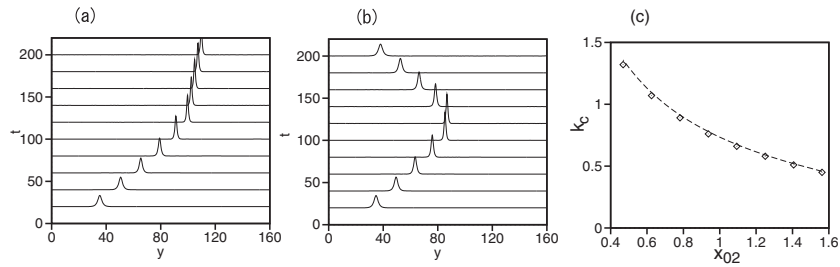}
\end{center}
\caption{ (a) Time evolution of $|\phi(x,y,t)|$ in the section of $x=0$ at $k=0.77$. (b) Time evolution of $|\phi(x,y,t)|$ in the section of $x=0$ fat $k=0.74$. (c) Critical value $k_c$ for reflection. The dashed curve denotes the critical curve estimated by the variational method.}
\label{f2}
\end{figure}
The approximate values of $a=1.357, b=1.353$, and $A=0.876$ are obtained by the variational method for $N=5, U_0=5$, and $x_0=3$. 
The dashed curve in Fig.~1(b) denotes $\phi=A{\rm sech}\{(y-\xi)/a\}$ where $\xi=20$, and the dashed curve in Fig.~1(c) denotes $\phi=A\exp\{-x^2/(2b^2)\}$. 
The variational method is a good approximation for the two-dimensional soliton in the channel.

Next, we study a problem in which the channel width changes in the $y$-direction as one of the simplest inhomogeneous systems.
We assume that the channel width $x_0(y)$ becomes narrow as $x_0(y)=x_{01}$ for $y<y_1$,  $x_0(y)=x_{01}+(x_{02}-x_{01})\times(y-y_1) /(y_2-y_1)$ for $y_1<y<y_2$, and $x_0(y)=x_{02}$ for $y>y_2$. The center line of the channel is fixed to $x=0$. We have performed direct numerical simulation from the initial condition $\phi(x,y)=\phi_0(x,y)\exp(ik y)$, where $\phi_0(x,y)$ is  the stationary solution shown in Fig.~1 for $N=5,x_0=3$, and $U_0=5$, and $k$ is the initial wave number in the $y$ direction. If the channel width is constant, the two-dimensional soliton propagates with the velocity $v_y=k$. 
The channel width changes smoothly, since $y_2-y_1$ is much larger than $x_{01}-x_{02}$.
The numerical simulation was performed by the pseudospectral method with $128\times 1024$ Fourier modes. The system size was $20\times 160$.
Figure 2(a) shows the time evolution for $k=0.77,x_{01}=3,x_{02}=2,y_1=40$, and $y_2=100$. The initial peak position of the soliton is $(0,20)$. The soliton  propagates into the narrower channel, but the velocity becomes slower and the soliton amplitude becomes higher in the narrower channel. Figure 2(b) shows the time evolution at $k=0.74$  for the same channel. The soliton cannot penetrate into the narrow channel and it is reflected. The threshold wave number $k_c$ is evaluated as $k_c=0.76$ for the channel. We have investigated the threshold value $k_c$ for various values of $x_{02}$ for $x_{01}=3,U_0=5,y_1=40$ and $y_2=100$. The rhombi in Fig.~2(c) show the relation of $k_c$ vs $x_{02}$ by direct numerical simulation. That is, $k_c$ decreases with $x_{02}$. 

\begin{figure}[bp]
\begin{center}
\includegraphics[height=3.5cm]{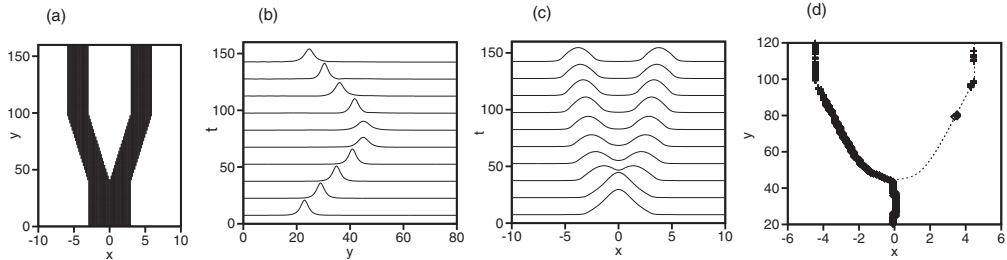}
\end{center}
\caption{ (a) Branching channel with $x_{01}=3,x_{03}=3,x_{02}=6,y_1=40$, and $y_2=100$.  (b) Time evolution of $|\phi(x,y,t)|$ at the section of $x=0$ for $k=0.4$. (c) Time evolution of $|\phi(x,y,t)|$ at the $x$-section passing through the peak position $(x_p,y_p)$ for $k=0.8$. (d) Trajectory of the peak position $(x_p(t),y_p(t))$ at $k=0.8$. The dashed curve is obtained by the variational method.}
\label{f3}
\end{figure}
\begin{figure}[bp]
\begin{center}
\includegraphics[height=3.5cm]{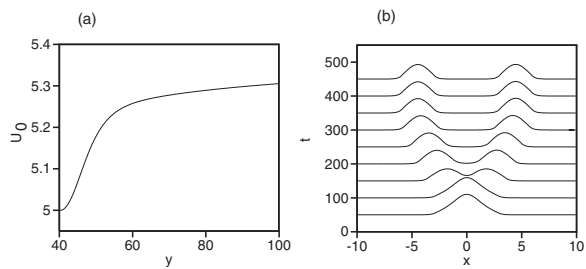}
\end{center}
\caption{ (a)  Depth of potential designed for no reflection by the variational method.   (b) Time evolution of $|\phi(x,y,t)|$ for the designed potential in the $x$-section passing through the peak position at $k=0.2$. (c) Time evolution of $|\phi(x,y,t)|$ for the designed potential at $k=0.1$ in the $x$-section passing through the peak position.}
\label{f4}
\end{figure}
To understand this phenomenon, we performed a variational method by assuming $\phi(x,y,t)=A{\rm sech}\{(y-\xi(t))/a\}\exp\{-x^2/(2b^2)\}\exp\{ip(y-\xi(t))-i\mu t\}$. For the assumption, the effective Lagrangian is expressed as
\begin{equation}
L_{eff}=N\{p\xi_t(t)-p^2/2-\frac{1}{4b^2}-\frac{1}{6a^2}+\frac{N}{6\sqrt{2\pi}ab}+U_0{\rm erf}(x_0/b)\}.
\end{equation}
The variation principle $d/dt(\partial L_{eff}/\partial \xi_t)=\partial L_{eff}/\partial \xi$ and $\partial L_{eff}/\partial p=0$ yield 
\begin{equation}
\frac{d\xi(t)}{dt}=p,\;\;\frac{dp}{dt}=-\frac{\partial U_{eff}}{\partial \xi},
\end{equation}
where $U_{eff}=1/(4b^2)+1/(6a^2)-N/(6\sqrt{2\pi}ab)-U_0{\rm erf}(x_0/b)$. The parameters $a$ and $b$ are evaluated using the variational principle at each point $y=\xi(t)$. 
Equation (7) is equivalent to  Newton's equation of motion with the effective potential $U_{eff}(y)$, which increases monotonically between $y_1$ and $y_2$. The threshold value $k_c$ of the wave number $k$ is evaluated from $k_c^2/2=U_{eff}(x_{02})-U_{eff}(x_{01})$. The dashed curve in Fig.~2(c) shows the threshold value evaluated by the variational method. The agreement with the direct numerical simulation is fairly good, but the theoretical curve is slightly larger than the numerically obtained values. 

The main problem in this study is a problem in which one channel branches into two channels.  The potential $U(x,y)$ is assumed to be $U(x,y)=-U_0$ for $-x_{01}<x<x_{01}$ when $y<y_1$, $U(x,y)=-U_0$ for $-x_{12}<x<-x_{13}$ and $x_{13}<x<x_{12}$,  $U(x,y)=-U_1$ for $-x_{13}<x<x_{13}$  when $y_1<y<y_1$, and $U(x,y)=-U_0$  for $-x_{02}<x<-x_{03}$ and $x_{03}<x<x_{02}$ when $y>y_2$, where $x_{12}=x_{01}+(x_{02}-x_{01})\times (y-y_1)/(y_2-y_1)$, $x_{13}=x_{03}\times(y-y_1)/(y_2-y_1)$, and $U_1=U_0\times(y_2-y)/(y_2-y_1)$. $U(x,y)=0$ in the other region. 
We introduced the $U_1$ component in the potential $U(x,y)$ for the potential to change smoothly in the $y$-direction.  
Figure 3(a) shows the channel region with $U(x,y)=-U_0$ for $x_{01}=3,x_{03}=3,x_{02}=6,y_1=40$, and $y_2=100$. 
The initial condition is the same as before:  $\phi(x,y,t)\sim A{\rm sech}\{(y-20)/a\}\exp\{-x^2/(2b^2)\}\exp\{ik(y-20)\}$. 
Figure 3(b) shows the time evolution of $|\phi(0,y,t)|$ in the section of $x=0$ at $k=0.4$. The other parameters are $U_0=5$ and $N=5$.
The soliton is reflected at the branching point. Figure 3(c) shows the time evolution of $|\phi(x,y_p)|$ in the $y$-section including the peak position of the modulus $|\phi|$ at $k=0.8$. The soliton with a single peak is split into two solitons propagating along the two channels. 
We investigated the threshold value $k_c$ for the transition from reflection to splitting, and obtained $k_c\sim 0.74$. 
Figure 3(d) shows the trajectory at the peak position for $k=0.8$ in the $(x,y)$ space. The peak position is located at $x=0$ for $y<43.8$ and it is split into two peaks for $y>43.8$. The peak positions continuously separate away from $x=0$. 

For the variational method applied to this splitting process, $\phi$ is assumed to be
$\phi=(A/2){\rm sech}\{(y-\xi(t))/a\}[\exp\{-(x-\eta)^2/(2b^2)\}+\exp\{-(x+\eta)^2/(2b^2)\}]\exp\{ip(y-\xi(t))-i\mu t\}$.
The variational principle yields, for this ansatz,
\begin{equation}  
\frac{d\xi(t)}{dt}=p,\;\;\frac{dp}{dt}=-\frac{\partial U_{eff}}{\partial \xi},
\end{equation}
where
\begin{eqnarray} 
U_{eff}&=&\frac{1}{6a^2}+\frac{1}{4b^2}-\frac{\eta e^{-\eta^2/b^2}}{2b^4(1+e^{-\eta^2/b^2})}-\frac{N}{12\sqrt{2\pi}ab}\frac{1+3e^{-2\eta^2/b^2}+4e^{-3\eta^2/(2b^2)}}{(1+e^{-\eta^2/b^2})^2},\nonumber\\
& &-U_0\frac{{\rm erf}\{(x_2-\eta)/b\}+{\rm erf}\{(x_2+\eta)/b\}+{\rm erf}\{(\eta-x_1)/b\}-{\rm erf}\{(\eta+x_1)/b\}}{2(1+e^{-\eta^2/b^2})},\nonumber\\
& &-U_0\frac{2e^{-\eta^2/b^2}\{{\rm erf}(x_2/b)-{\rm erf}(x_1/b)\}}{2(1+e^{-\eta^2/b^2})},\nonumber\\
& &-U_1\frac{{\rm erf}\{(x_1+\eta)/b\}-{\rm erf}\{(\eta-x_1)/b\}+2e^{-\eta^2/b^2}{\rm erf}(x_1/b)}{2(1+e^{-\eta^2/b^2})},
\end{eqnarray}
where $x_1=0,x_2=x_{01}$ for $y<y_1$, $x_1=x_{13}$, and $x_2=x_{12}$ for $y_1<y<y_2$, and $x_1=x_{03}$, and $x_2=x_{02}$ for $y>y_2$.
In Fig.~3(d), the peak positions evaluated by the variational method are drawn with dashed curves. The trajectory is well approximated by the variational method. The threshold value obtained by the variational method is $k_c=0.77$, which is comparable to the numerical one, $k_c\sim 0.74$, but the theoretical estimate is slightly larger than the numerical one, which is similar to the case of the tapering channel.   

The reflection of the pulse at the branching point might be undesirable for applications to pulse splitting. We can design a system including a branching point with reduced reflection using the variational method. The difference $U_{eff}(y_2)-U_{eff}(y_1)$ of the effective potential was the origin of the reflection. The reflection is expected to disappear or be reduced, if the channel system is designed, in which the difference $|U_{eff}(y_2)-U_{eff}(y_1)|$ is minimized. 
We can control both the channel width $x_{02}-x_{03}$ and the depth of potential $U_0(y)$. We have designed the value of $U_0(y)$ using an evolution equation: $dU_0(y)/d\tau=-\beta(U_{eff}(y_1)-U_{eff}(y))$ with $\beta>0$ as $U_{eff}(y)$ approaches $U_{eff}(y_1)$ at each point $y>y_1$. The $U_1$ component is assumed to be the same as before: $U_1(y)=5\times(y_2-y)/(y_2-y_1)$. Figure 4(a) shows the designed value of $U_0(y)$ for  $x_{01}=3,x_{02}=2,y_1=40$, and $y_2=100$. Figure 4(b) shows the time evolution of the soliton with the initial wave number $k=0.2$. Even for this small wave number, the reflection is suppressed and the soliton is split into two solitons, that penetrate into the two channels. 

In summary, we have studied the motion of two-dimensional solitons confined in guided channels. We have found that solitons with small wave number are reflected at the branching point of the channel. The variational method can predict the reflection and splitting phenomena fairly well. The method can be applied in the design of channel systems with reduced reflection.

\end{document}